# Observation of non-local effects in ion transport channel in J-TEXT plasmas

Yuejiang Shi[1*], Zhoujun Yang[2*], Zhongyong Chen[2], Zhifeng Cheng[2], Xiaoyi Zhang[2], Wei Yan[2], Jie Wen[5], Qinxue Cai[2], Kaijun Zhao[3], Seulchan Hong[1], JaeMin Kwon[4], Partick H Diamond[3,6], Peng Shi[2], Hao Zhou[2], Xiaoming Pan[2], Zhipeng Chen[2], SeongMoo Yang[1], Yunbo Dong[5], Lu Wang[2], YongHua Ding[2], Yunfeng Liang[2,7], Zhongbin Shi[5], Yong-Su Na[1]

[1]Department of Nuclear Engineering, Seoul National University, Seoul, Korea

[2] International Joint Research Laboratory of Magnetic Confinement Fusion and Plasma Physics, State Key Laboratory of Advanced Electromagnetic Engineering and Technology, School of Electrical and Electronic Engineering, Huazhong University of Science and Technology, Wuhan, China

[3]College of Nuclear Science and Engineer, East China University of Technology, Nanchang, China

[4]National Fusion Research Institute, Daejeon, Korea

[5]Southwestern Institute of Physics, Chengdu, China

[6]CMTFO and CASS, University of California, San Diego, USA

[7]Forschungszentrum Jülich GmbH, Institut für Energie-und Klimaforschung-Plasmaphysik, Partner of the Trilateral Euregio Cluster (TEC), Jülich, Germany

[*]E-mail of corresponding author: yjshi@ipp.ac.cn and yangzj@hust.edu.cn

**Abstract**. In cold pulse experiments in J-TEXT, the ion transport shows similar non-local response as the electron transport channel. Very fast ion temperatures decreases are observed in the edge, while the ion temperature in core promptly begin to rise after the injection of cold pulse. Moreover, the cutoff density is also found for the ion non-local effect. The experimental observed density fluctuation in a high frequency ranging from 500 kHz to 2 MHz is obviously reduced in the whole plasma region during non-local transport (NLT) phase.

There are several outstanding experimental mysteries in magnetic fusion plasma research. A fast increase of the central electron temperature caused by the edge cooling in Ohmic heating (OH) plasmas, the so-called non-local heat transport (NLT) [1] is one of these challenging issues. NLT was first observed in the TEXT tokamak 23 years ago, and in many fusion devices afterwards (TFTR[2], RTP[3], ASDEX UPGRAGE[4], Tore Supra[5], JET[6], LHD[7-9], HL-2A[10-12], Alcator C-Mod[13-15], KSTAR[16], J-TEXT[17], and EAST[18]). NLT phenomenon shows that the limitation of transport theory based on traditional local transport model. The underlying physics mechanism of NLT is still unclear.

In a nutshell, the NLT phenomena in cold pulse experiment were only reported and focused on electron heat transport channel in most of the previous experimental research. However, our research results in J-TEXT show that NLT effects also exist in particle transport channel [17]. The rotation or momentum transport also has some possible NLT response in cold pulse experiment [17]. Recently, the temporal resolution of the diagnostic systems for ion temperature measurement in J-TEXT is upgraded to millisecond level, which provides us the capability to investigate the dynamic response of ion transport channel in cold pulse experiment. The latest experimental results in J-TEXT tokamak in this paper show that NLT effects are exist in ion heat channel indeed. The edge ion temperature decreases promptly after injection of cold pulse, which is accompanied by rapid core ion temperature increases. Moreover, the increases of core ion temperature disappear at high density cold pulse plasma. These new experimental findings of non-local transport in J-TEXT in this paper provide a key clue to reveal more clear background physics mechanism for NLT. All the results presented in this paper are obtained from J-TEXT tokamak [19].

The major radius of J-TEXT is 1.05m. The minor radius of J-TEXT is 0.255m in this experiment. All the discharges are OH heating plasmas with circular limiter configuration. In J-TEXT, the electron temperature ($T_e$) is measured with the electron cyclotron emission radiometer (ECE) [20]. A 17-channel polarizer-interferometry system (POLARIS) [21] which covers the whole region of J-TEXT plasma from high field side to low field side, measures the profiles of electron density ($n_e$). The x-ray crystal spectrometer (XCS) [22] provides the ion temperature ($T_i$) of core plasma. A fast passive visible spectroscopy (ERD) system [23] can provide the ion temperature and toroidal rotation velocity ($V_\phi$) at two edge radii (r/a~0.7 and r/a~0.93). At present, the fast ERD cannot simultaneously measure the two positions during one shot. Shot by shot scan is applied to obtain the data for two edge channels. In the latest experimental campaign in J-TEXT, the temporal resolution of ERD with improved optical system and XCS with new detector can reach 1ms and 2ms, respectively. The density fluctuations and poloidal rotation velocity in some discharges are measured with Doppler Backscattering reflectometer (DBS). Multi-pulse supersonic molecular beam injection (SMBI) [24] is applied as cold pulse source to trigger NLT in J-TEXT [25]. The toroidal magnetic field ($B_T$) in our experiments is fixed at 1.8T, which make the best spatial measured coverage of ECE.

Fig.1 shows the waveforms of two representative low density OH shots with five SMBI cold pulse injection in J-TEXT. As mentioned above, full experimental data cannot be obtained in single shot. The discharge parameters for the two shots in fig.1 are identical, which make all of the experimental data effective for the investigation of one physics phenomena. The pulse duration of SMBI is 0.3ms. The time interval between each SMBI pulse is 50ms. The plasma current ($I_p$) and edge safety factor ($q_a$) are 150kA and about 3.7, respectively. The typical

NLT effect in electron heat transport channel (core $T_e$ rises while edge $T_e$ drops) appears in every SMBI pulse, which is same as previous NLT experiments in other devices [1-12]. On the other hand, the prompt increasing of core ion temperature and rapid decreases of edge $T_i$ due to SMBI are also clearly shown in fig.1. Obviously, the responses of the ion temperature to the cold pulse have the same NLT characters as the electron temperature. As shown in fig.1, the decreases of edge toroidal rotation are also clear and remarkable.

One typical characteristic of NLT in electron channel is that the increments of core $T_e$ temperature due to injection of cold pulse will not appear in high density plasma. This critical density where the electron NLT effect disappears is defined as the cutoff density. Fig.2 shows the relation between increments of core temperature induced by cold pulse and the line average density. It can be seen that the core $T_i$ rise in response to cold pulse disappears at high plasma density. The response of core $T_i$ in high density plasma cold pulse experiments show the same trend as that of core $T_e$.

Up to now, many theoretical models or interpretations have been proposed to explain NLT phenomenon [2-4, 6, 9, 11-14, 26-35]. In general, there are two strategies to interpret the NLT. One is the based on the fast change of confinement properties during NLT phase. The other one considers the increased heating power or additional heating source during cold pulse injection. Fig.3 shows the profiles of density and temperature for the second SMBI pulse in fig.1.The electron heat diffusion coefficient $\chi_e$ for J-TEXT's NLT plasma has been calculated from power balance and shown in fig. 4. The calculation of $\chi_e$ is based on the assumption that the heating power is constant during SMBI injection phase. It can be seen that the $\chi_e$ drops in a wide region from edge to core. Similar to the previous NLT results in other devices [], the rapid drop in the $\chi_e$ profile can explain the experimental results. At the same time, the density fluctuation in a high frequency ranging from 500 kHz to 2 MHz measured from DBS in fig.5 clearly shows the turbulence suppression during cold pulses injection phase. The reduction of the density fluctuation can be observed in the whole main plasma region, which is consistent with the drops of $\chi_e$ in a wide region fig.4. Moreover, the time scale of the density fluctuation transition also matches with the fast increase of core temperature. These experimental evidences suggest that the confinement improvement due to suppression of turbulence maybe the main mechanism for NLT phenomenon.

However, why confinement has such transient change is more fundamental and important to understand some basic transport characters of fusion plasma. $E \times B$ flow shear stabilization is the most important and successful model to explain the turbulence suppression and confinement improvement. The poloidal rotation profiles from DBS for J-TEXT's NLT plasma are shown in fig.6, which clear shows that both flows and flow shear are reduced after the injection of SMBI. On the other hand, the ion temperature gradient (ITG) micro-instability model [29, 30] has also been proposed to interpret NLT. The main idea of the ITG-based mode is that both the ion and electron effective diffusivities are affected by the ITG-driven turbulence or instability. The response of $T_i$ is the key issue in the ITG-based model because the ITG instability threshold increases with ration of $T_i/T_e$. The simulation results of the ITG-based mode quantitatively produce the experimental core $T_e$ rise in TEXT [29, 30]. Moreover, prompt core $T_i$ rise is also predicted by the ITG-based mode. Now the experimental data in J-TEXT verify the core $T_i$ rise by cold pulse indeed. However, the core $T_e$ or $T_i$ rise in the ITG-based mode is based on the assumption of edge $T_i$ rise. On the contrary, the experimental data in this paper show edge $T_i$ decrease due to cold pulse. On the other hand, the gyro-kinetic simulation results for the cold pulse experiment with strong electron cyclotron heating (ECH) in KSTAR [16] clearly show the trapped electron mode (TEM) is the dominant turbulence. It is obvious that the mode based on pure ITG has some fundamental defects to explain the NLT phenomenon. It is no doubt that supplement of TEM to ITG-based mode is more reasonable solution. Recently, the improved ITG/TEM-based mode has applied to interrupt the experimental NLT data in Alcator C-mod [35]. However, the edge $T_i$ still increases and core $T_i$ decreases after injection of cold pulse in the ITG/TEM-based mode, which is completely contrary to the experimental response of $T_i$ in J-TEXT.

On the other hand, since ITG/TEM-based model can be classified to theoretic category of micro-instability or turbulence, we carry out linear micro-instability analysis for J-TEXT's NLT plasmas with GKW code [36]. The analysis results of GWK simulation for these plasma parameters in fig.3 are shown in fig.7. It can be seen in fig.7 that the growth rates of relative low wavenumber turbulence ($k_\theta\rho_s<1$) are similar for plasma before SMBI injection and plasma after SMBI injection. Compared to the plasmas before SMBI injection, the growth rate at $k_\theta\rho_s=0.5$ decrease about 9% for plasma at r/a=0.3 and increase about 8% for plasma at r/a=0.6. On the other hand, the change of the growth rate of relative high wavenumber turbulence ($k_\theta\rho_s>1$) is clear and obvious. The growth rates at $k_\theta\rho_s>1$ for all the regions from r/a=0.3 to r/a=0.6 dramatically increase for the plasmas after SMBI injection. Compare to the change of turbulence at high wavenumber, the change of turbulence at low wavenumber can be neglected. Overall, it is very difficult to find the clear evidence of turbulence suppression during NLT from the simulation results based on the critical gradient mode for J-TEXT's plasmas.

Turbulence spreading (TS) model is also one remarkable theoretical hypothesis to explain NLT phenomenon. The detail mechanism of TS mode can be seen in ref [33, 34]. From the views of experimental data, the global reduction of flow shear as shown in fig.6 can boost the large scale avalanche transport from edge to core through turbulence spreading. At the same time, the small time delay for the occurrence of fluctuation reduction between the edge channel and core channel as shown in fig.5 coincides with the fast propagation property of turbulence spreading. Moreover, TS model is only one mode to clearly predict the response of toroidal rotation in cold pulse experiment. Both core rotation and edge rotation will increase by cold pulse in TS mode [34]. Some indirect evidence of acceleration of core rotation has been found in J-TEXT [17]. However, the directly measurement results in this paper clearly show the drop of edge rotation.

As described above, the multi-channel (electron, ion, particle, and momentum) NLT phenomenon in J-TEXT is not fully consistent with the prediction of the confinement improvement theoretical models based on $E \times B$ flow shear stabilization or instability related to critical gradient. On the other hand, additional heating is one possible solution. As Pustovitov pointed out in ref [37], the magnetic field surrounding the plasma can provide the necessary heating source to make NLT. There is energy transfer between the plasma and the magnetic field when plasma edge is cooling [37]. In other words, the magnetic field heats the plasma by radial compression during cold pulse injection. The simultaneous increment of electron temperature, ion temperature, and density in core region can be achieved with this heating mechanism. On the other hand, the contribution of confinement improvement mechanism cannot be neglected because the experimental data clearly shows the suppression of high frequency density fluctuation during NLT. Essentially all modes are wrong, but some are useful [38]. These new experimental events observed in J-TEXT are a real challenge to the current understanding of the non-local transport phenomenon. On the other hand, more new experimental data can provide more opportunities for the improvement of current transport models or proposal of some new ideas. The investigation about the background mechanism for NLT plasma is still an open question and worth continuing to be explored in future research.

**Acknowledgement**

This research is supported by Basic Science Research Program through the National Research Foundation (NRF) funded by the Ministry of Science and ICT of the Republic of Korea (No. 2018R1A2B2008692 and No.2014M1A7A1A03045368). This research is also supported by National Magnetic Confinement Fusion Science Program funded by Ministry of Science and Technology of China (No. 2014GB108001, 2015GB111002, and 2015GB120003) and National Natural Science Foundation (NSFC) of China (No. 11775089).

**Reference**

[1] K. W. Gentle *et al*, Phys. Rev. Lett. **74**, 3620 (1995).

[2] M. W. Kissick *et al*, Nucl. Fusion **36,** 1691 (1996).

[3] P. Mantica *et al*, Phys. Rev. Lett. **82**, 5048 (1999). (Shell model)

[4] F. Ryter *et al*, Nucl. Fusion **40** , 1917 (2000). (ITG explaination)

[5] X. L. Zou *et al*, Plasma Phys. Control. Fusion **42**, 1067 (2000).

[6] P. Mantica *et al*, Plasma Phys. Control. Fusion **44**, 2185 (2002).

[7] N. Tamura *et al*, Phys. Plasmas **12**, 110705 (2005).

[8] S. Inagaki *et al*, Nucl. Fusion **46**, 133 (2006).

[9] K. Ida *et al*, Nucl. Fusion **55**, 013022 (2015).

[10] H. J. Sun *et al*, Plasma Phys. Control. Fusion **52**, 045003 (2010).

[11] H. J. Sun *et al*, Nucl. Fusion **51**, 113010 (2011).

[12] O.Pan *et al*, Nucl. Fusion **55**, 113010 (2015).

[13] J. E. Rice *et al*, Nucl. Fusion **53**, 033004 (2013).

[14] C. Gao *et al*, Nucl. Fusion **54**, 083025 (2014).

[15] P. Rodriguez-Fernandez *et al*, Nucl. Fusion **57**, 074001 (2017).

[16] Y.J.Shi *et al*, Nucl. Fusion.**57**, 066040 (2017).

[17] Y.J.Shi *et al*, Nucl. Fusion. **58**, 044002 (2018).

[18] Y.Liu and Y.J.Shi , *et al*, Nucl. Fusion **59**, 044005 (2019).

[19] G. Zhuang *et al*, Nucl. Fusion **51**, 094020 (2011).

[20] Z.J.Yang *et al*, Rev. Sci. Instru*m.* **87**, 11E112 (2016).

[21] J.Chen *et al*, Rev. Sci. Instrum. **85**, 11D303 (2014).

[22] W.Yan *et al*, Rev. Sci. Instrum. **87**, 11E318 (2016).

[23] Z.F.Cheng *et al*, Rev. Sci. Instrum. **84**, 073508 (2013).

[24] L.H.Yao *et al*, Nucl. Fusion **38**, 631 (1998).

[25] J.S.Xiao *et al*, IEEE TRANSACTIONS ON PLASMA SCIENCE **41**, 3675(2013)

[26] P.H. Diamond and T.S.Hahm, *Phys. Plasmas* **2,** 3640(1995).

[27] V.V. Parail *et al,* Nucl. Fusion **37**, 481 (1997)

[28] J.D.Callen and M.W.Kissik, Plasma Phys. Control. Fusion **39**, B173(1997)

[29] J.E. Kinsey *et al,* 1998 *Phys. Plasmas* **5** 3974 (1998)
[30] J.E. Kinsey *et al,* 1999 *Phys. Plasmas* **6** 1865 (1999)
[31]A. K. Wang *et al*, Nucl. Fusion **49**, 075025 (2009).

[32] X.Q.Ji *et al*, Sci. Rep. **6**, 32697 (2016).

[33] V. Naulin *et al*, Rotation reversal in a 1D turbulence spreading model *Proceedings of the 41st EPS Conference on Plasma Physics: Europhysics Conference* (Berlin, Germany, 23-27 June 2014)Vol. **38F** P2.067, http://ocs.ciemat.es/EPS2014PAP/pdf/P2.067.pdf

[34] F. Hariri *et al*, Phys. Plasmas **23**, 052512 (2016).

[35] P. Rodriguez-Fernandez *et al*, Phys. Rev. Lett. **120**, 075001(2018).

[36] A.G. Peeters *et al,* Comput. Phys. Commun. **180**, 2650 (2009)

[37] V.D. Pustovitov, Plasma Phys. Control. Fusion **54**, 124036(2012)

[38] G. E. P. Box, Journal of the American Statistical Association **71**, 791(1976)

**Figures**

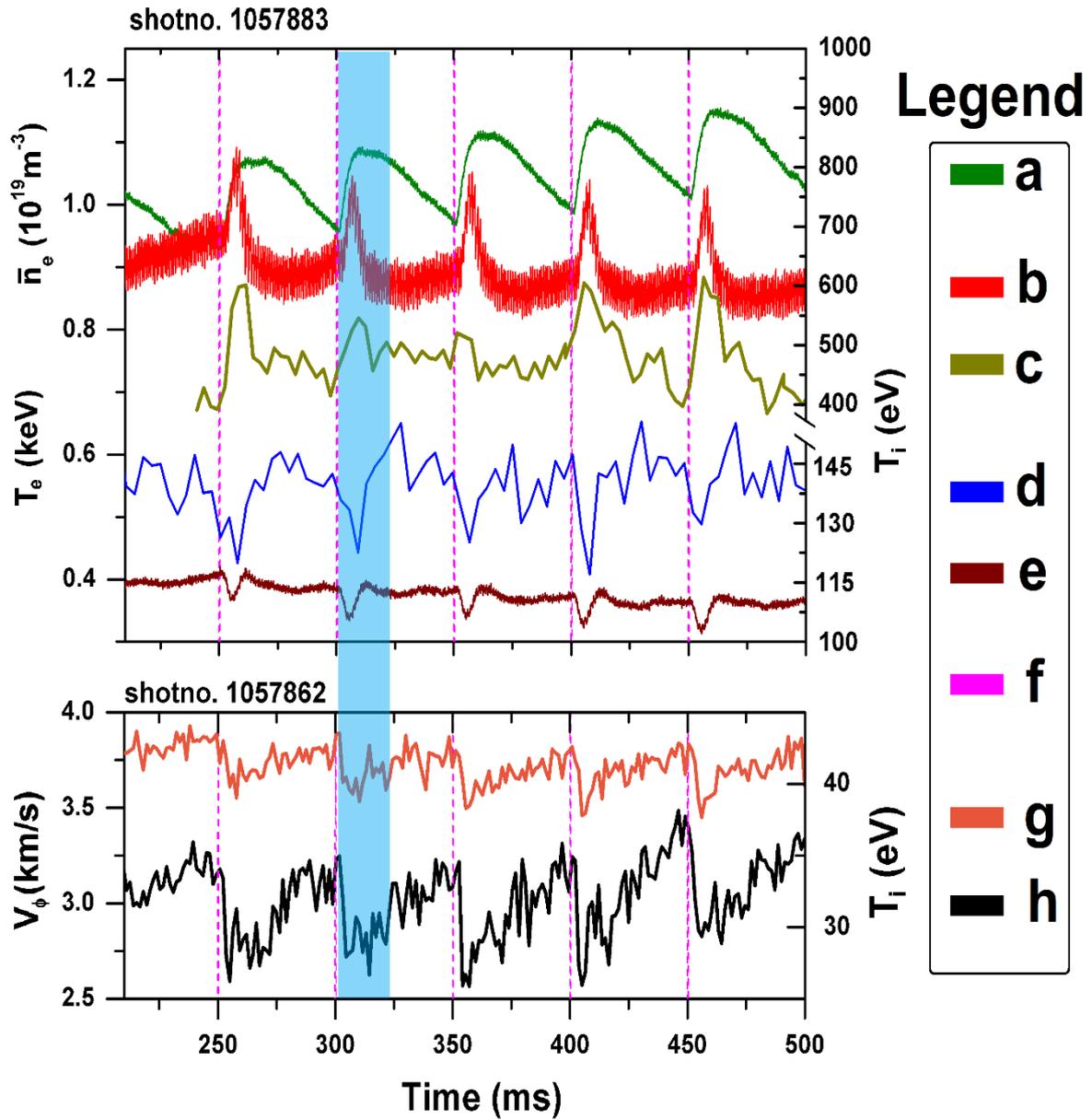

*Fig.1* The waveforms of cold pulse discharge with multi-pulse SMBI (shot no.1057883 and 1057862). (a) line-averaged electron density; (b) core electron temperature; (c) core ion temperature; (d) edge ion temperature at $r/a\sim0.7$; (e) edge electron temperature at $r/a=0.69$; (f) SMBI pulse signal; (g) edge ion temperature at $r/a\sim0.93$; (h) edge toroidal rotation velocity at $r/a\sim0.93$.

2019NLT-V3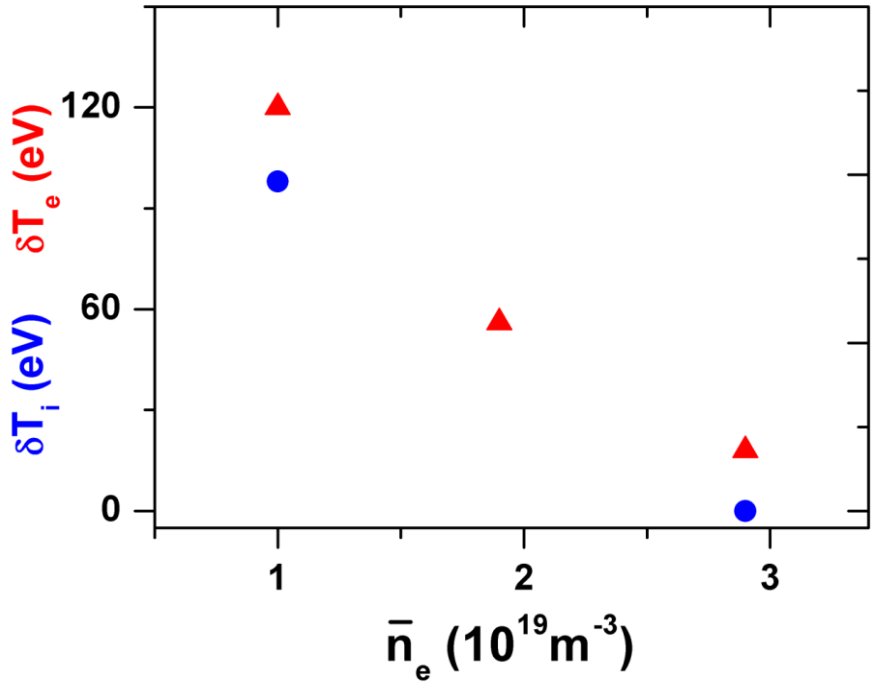

*Fig.2 The relation between the centre temperature rise due to NLT effect and line averaged density for the plasmas with same operating parameters ($B_T$=1.8T, $I_p$=150kA).*

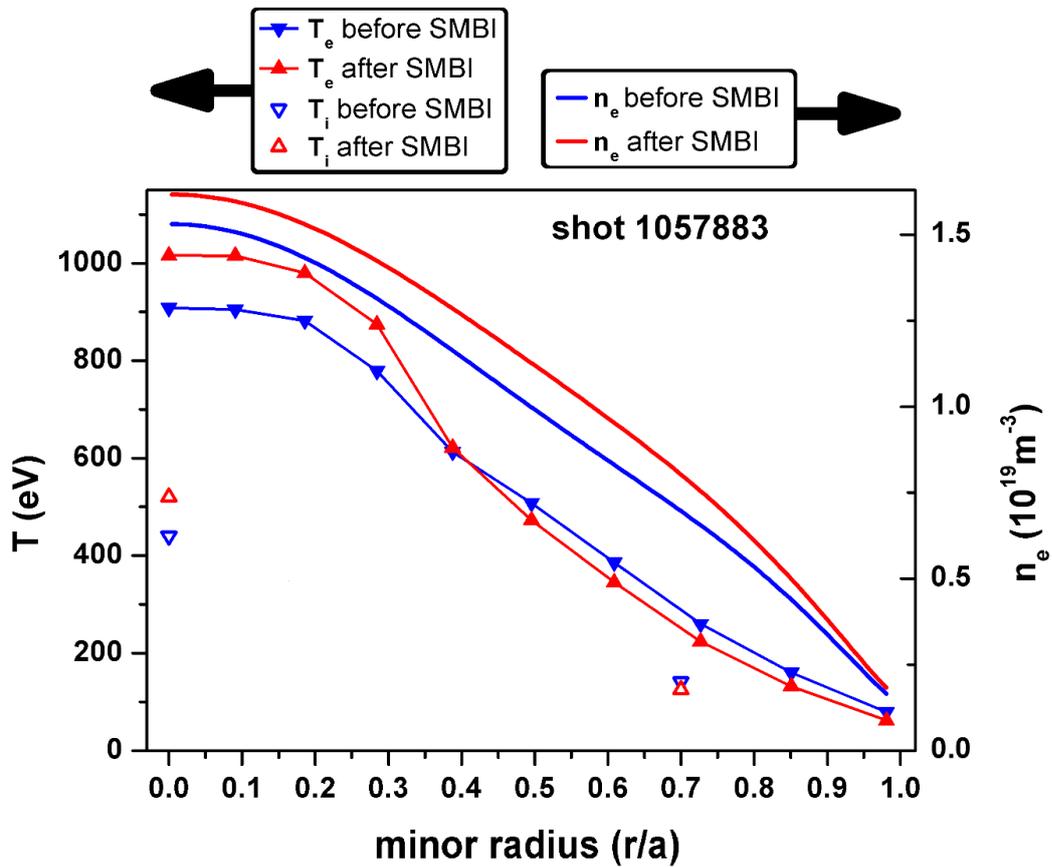

*Fig.3 The profiles of temperature and density for the seconds SMBI pulse in fig.1*

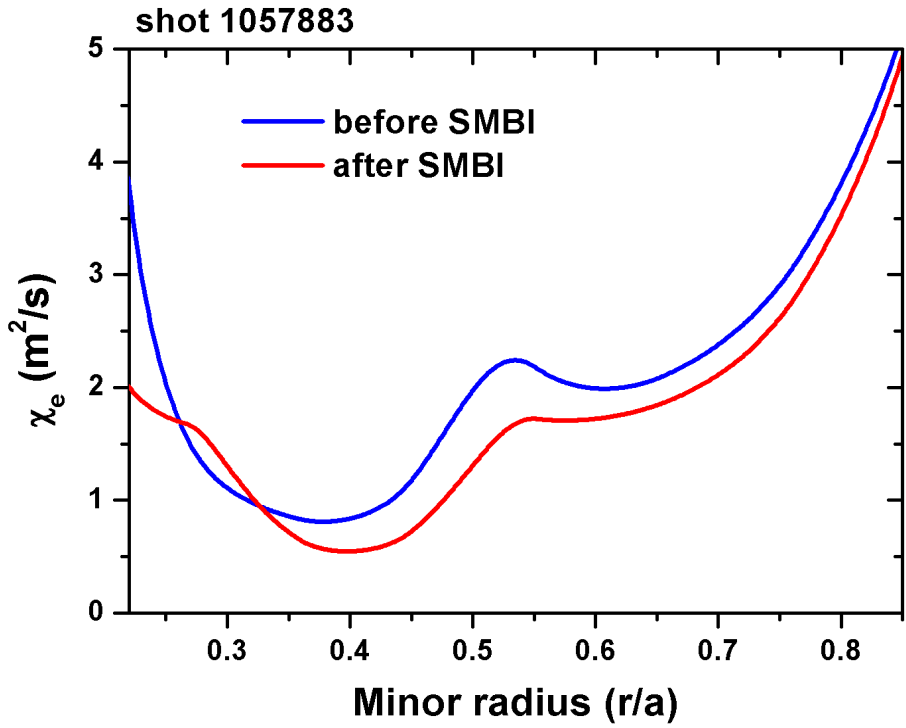

*Fig.4 The electron heat diffusivity profiles for J-TEXT's NLT plasmas.*

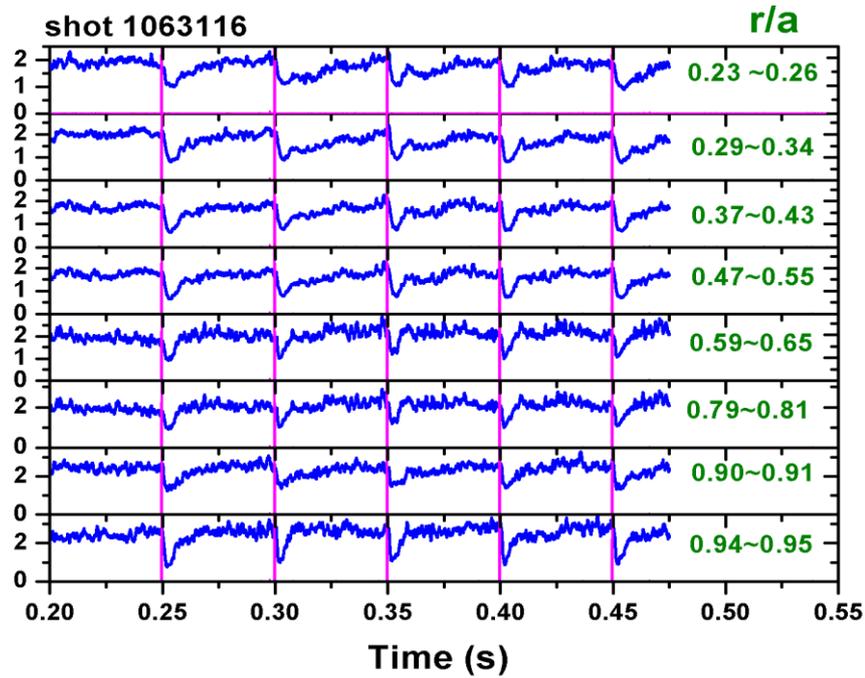

*Fig.5 Time trace of the density fluctuation power integrated over 500kHz-2MHz.*

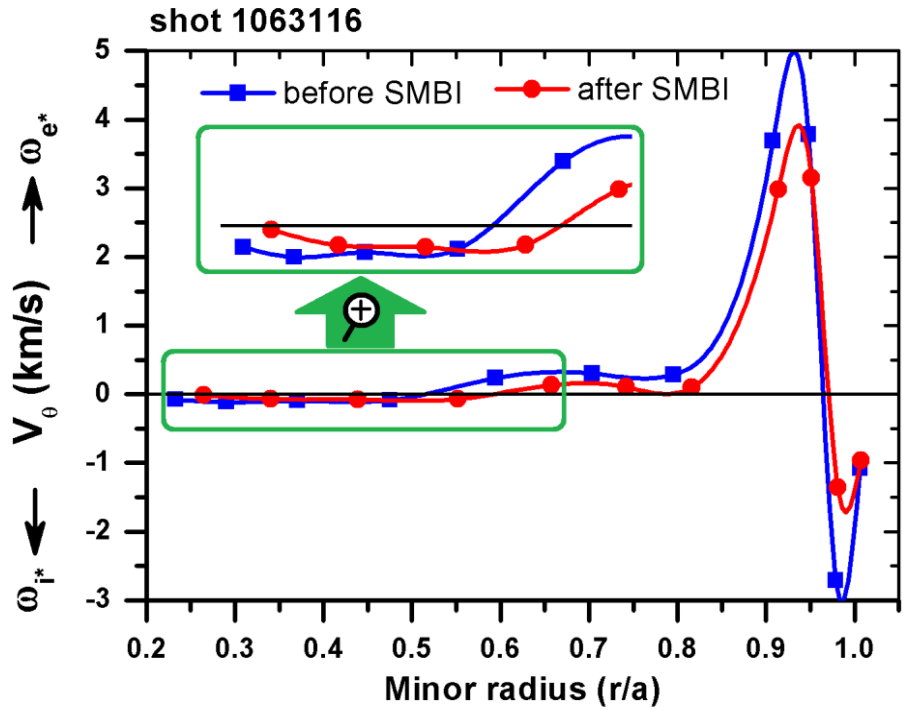

Fig.6 The poloidal flow profiles for J-TEXT's NLT plasmas

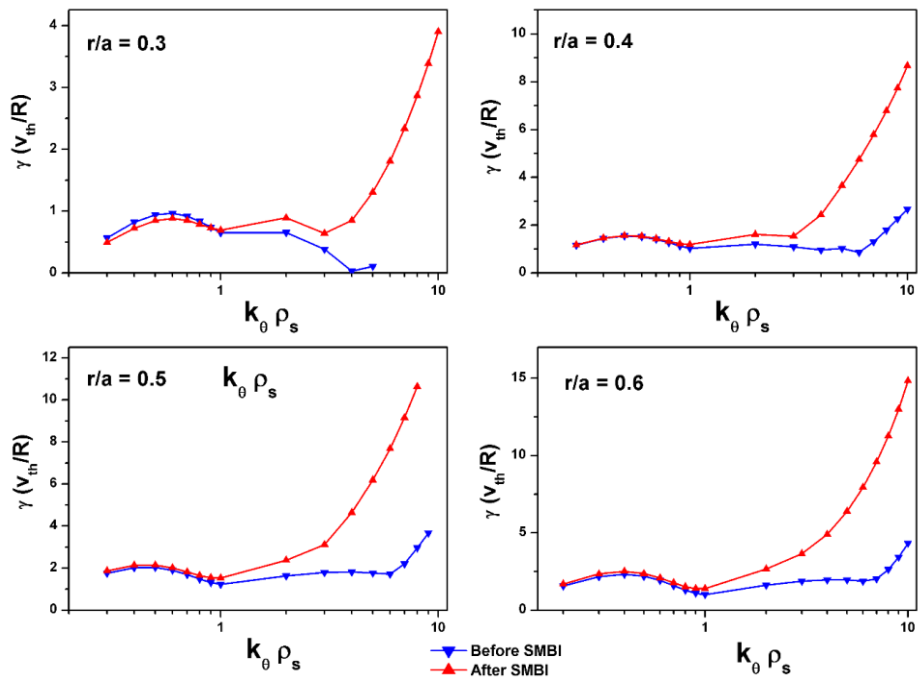

Fig.7 The linear growth rate simulated with GKW for the plasmas in fig.3.